\input harvmac
\input epsf

\overfullrule=0pt
\abovedisplayskip=12pt plus 3pt minus 3pt
\belowdisplayskip=12pt plus 3pt minus 3pt
%
\def\tilde{\widetilde}

\def\to{\rightarrow}
\def\tphi{{\tilde\phi}}
\font\zfont = cmss10 

\def\bigone{\hbox{1\kern -.23em {\rm l}}}
\def\ZZ{\hbox{\zfont Z\kern-.4emZ}}


\lref\pold{J. Polchinski, {\it ``D-branes and Ramond-Ramond charges''},
hep-th/9510017, Phys. Rev. Lett. {\bf 75} (1995), 47.}
\lref\witbound{E. Witten, {\it ``Bound states of strings and
p-branes''}, hep-th/9510135, Nucl. Phys. {\bf B460} (1996), 335.}
\lref\stromopen{A. Strominger, {\it ``Open p-branes''}, hep-th/9512059,
Phys. Lett. {\bf B383} (1996), 44.}
\lref\asy{O. Aharony, J. Sonnenschein and S. Yankielowicz, {\it
``Interactions of strings and D-branes from M theory''},
hep-th/9603009, Nucl. Phys. {\bf B474} (1996), 309.}
\lref\schwrev{J.H. Schwarz, {\it ``Lectures on superstring and M
theory dualities''}, hep-th/9607201, Nucl. Phys. Proc. Suppl. {\bf
55B} (1997), 1.}
\lref\gabzwie{M. Gaberdiel and B. Zwiebach, {\it ``Exceptional
groups from open strings''}, hep-th/9709013.}
\lref\dmf{K. Dasgupta and S. Mukhi, {\it ``F-theory at constant 
coupling''}, hep-th/9606044, Phys. Lett. {\bf B385} (1996), 125.}
\lref\calmal{C. Callan and J. Maldacena, {\it ``Brane dynamics from
the Born-Infeld action''}, hep-th/9708147.}
\lref\schwmult{J.H. Schwarz, {\it ``An SL(2,Z) multiplet of type IIB
superstrings''}, hep-th/9508143, Phys. Lett. {\bf B360} (1995), 13.}
\lref\chernsimons{
M. Li, {\it ``Boundary states of D-branes and Dy strings''},
hep-th/9510161,  Nucl. Phys. {\bf B460} (1996), 351\semi
M. Bershadsky, V. Sadov and C. Vafa, {\it ``D-branes and 
topological field theories''}, hep-th/9611222, Nucl. Phys. {\bf B463}
(1996), 420\semi
M.B. Green, J.A. Harvey and G. Moore, 
{\it ``I-brane inflow and anomalous couplings on D-branes''},
hep-th/9605033, Class. Quant. Grav. {\bf 14}(1997), 47.}
\lref\banks{T. Banks, {\it ``Matrix theory''}, hep-th/9710231, and
references therein.}
\lref\gibbons{G. Gibbons, {\it ``Born-Infeld particles and Dirichlet
p-branes''}, hep-th/9709027.} 
\lref\lpt{S. Lee, A. Peet and L. Thorlacius, {\it ``Brane-waves and
strings''}, hep-th/9710097.}
\lref\thor{L. Thorlacius, {\it ``Born-Infeld string as a boundary 
conformal field theory''}, hep-th/9710181.}

{\nopagenumbers
\Title{\vtop{\hbox{hep-th/9711094}
\hbox{TIFR/TH/97-58}}}
{\centerline{BPS Nature of 3-String Junctions}}
\centerline{Keshav Dasgupta\foot{E-mail: keshav@theory.tifr.res.in}
and Sunil Mukhi\foot{E-mail: mukhi@theory.tifr.res.in}}
\vskip 2pt
\centerline{\it Tata Institute of Fundamental Research,}
\centerline{\it Homi Bhabha Rd, Mumbai 400 005, India}
\ \smallskip
\centerline{ABSTRACT}

We study BPS-saturated classical solutions for the world-sheet theory
of a D-string in the presence of a point charge. These solutions are
interpreted as describing planar 3-string junctions, which arise
because the original D-string is deformed by the presence of the
inserted charge. We compute the angles of the junctions and show that
the vector sum of string tensions is zero, confirming a conjecture of
Schwarz that such configurations are BPS states.

\Date{November 1997}
\vfill\eject}
\ftno=0

D-branes\refs{\pold} are those extended objects on which an open
string can end. More precisely, a single D-brane is a state defined by
having both ends of an open string end on it. If a single end of an
open string ends on a D-brane (and the other end goes to infinity or
terminates on a different brane) then in principle we face a problem
with charge conservation: the charge of the open string appears to be
lost. The resolution to this\refs{\witbound,\stromopen}, which brought
about an important conceptual advance in the understanding of branes
in general, is that the world-volume theory of D-branes has certain
couplings which render this termination consistent.

As a result, the 2-form charge carried by the open string gets
converted into the gauge charge of the particle represented by its
endpoint. This charge can be measured by enclosing the endpoint by a
$(p-1)$-sphere within the $p$-brane. This has an important consequence
when $p=1$\refs{\asy,\schwrev}. In this case there is a discontinuity in
the charge measured on the 1-brane (D-string), so that it is not
possible for both ends of the D-string to remain D-strings. If we
denote a fundamental string (F-string) as a $(1,0)$ string, and a
D-string as a $(0,1)$ string, then when these two meet, the third
``outgoing'' string must be a $(1,1)$ string. In general, 3-string
junctions exist where three open string carrying charges
$(p_i,q_i),~i=1,2,3$ meet at a point, and charge conservation requires
$\sum_i p_i = \sum_i q_i =0$. (We assume that the three strings are
all oriented the same way with respect to the junction.)
\bigskip

\centerline{\epsfbox{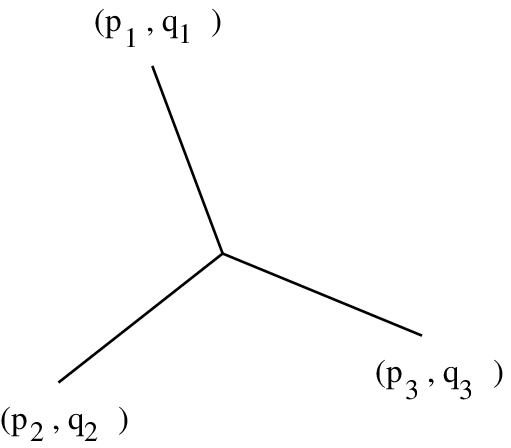}}
\centerline{Fig.1: A 3-string junction.}
\bigskip

It was conjectured in Ref.\refs{\schwrev} that such 3-string junctions
correspond to BPS saturated states in the theory, whenever the
configuration consists of straight strings going off to infinity from
the junction along definite angles. The angles are such that the
vector sum of string tensions is zero. These 3-string junctions
recently played an important role in Ref.\refs{\gabzwie} where they
were needed to explain the origin of exceptional symmetries in
F-theory, following the proposal of Ref.\refs{\dmf}. 

In what follows, we provide evidence for the existence of BPS
saturated 3-string junctions having precisely the properties described
above. Instead of looking for them as solutions of the classical
supergravity field equations, as suggested in Ref.\refs{\schwrev}, we
will find them by examining worldsheet properties of the D-string.

The worldsheet field theory for a single D-string is given, in the
linearized approximation, by a 2d supersymmetric $U(1)$ gauge theory.
The bosonic fields are the gauge field and 8 scalars representing
collective coordinates for the transverse motion of the D-string. Let
us now consider a simple case where a point charge of unit magnitude
is inserted onto a single D-string. This is a special case of the
situation studied in Ref.\refs{\calmal}, where point charges are
inserted onto general D $p$-branes. We take the D-string to be
oriented along the $x^1$ axis. Note that at present, the RR scalar
$\tphi$ is not excited.

In the presence of this charge, Gauss' law in one space dimension
states that 
\eqn\gausslaw{
F_{01}^+ - F_{01}^- = g }
where $g$ is the strength of the point charge, equal to the type IIB
string coupling. This requires the scalar potential to be piecewise
linear, with a fixed discontinuity in its slope. A simple solution of
this equation is
\eqn\simpsol{
\eqalign{
A_0 &= - g x^1,\quad x^1>0\cr
&=0,\qquad\qquad x^1<0\cr}}
This solution by itself is not BPS saturated but, as in Ref.{\calmal},
it can be made BPS by simultaneously exciting one of the transverse
coordinates, say $x^9$, represented by the worldsheet field
$X^9(x^0,x^1)$. Thus we choose
\eqn\xnine{
X^9(x^1)= A_0(x^1)}
with $A_0$ given by the above equation. 

Because of well-known properties of low-dimensional electrodynamics,
the solution is linearly increasing away from the charge, in sharp
contrast to the case for $p$-branes with $p\ge 3$. Indeed, the linear
variation of $X^9$ as a function of $x^1$ suggests the followng
interpretation: inserting the point charge at the origin of the
D-string causes one half of the string to rigidly bend. The point
charge itself, as in Ref.{\calmal}, is associated to the endpoint of
an F-string coming in perpendicular to the original D-string (in
contrast to the cases discussed in Ref.\refs{\calmal}, our solution
has no ``spike'' representing the F-string. However, a consistent
interpretation certainly requires the F-string to be present, as it is
the one which carries the inserted charge).

The resulting configuration is as in Fig 2, with the angle $\alpha$
determined by
\eqn\anglealpha{
\tan \alpha = {1\over g} }
The string that goes out from the junction towards the top left is
neither an F-string nor a D-string, but must be thought of as a
$(1,1)$ or $(-1,-1)$ string depending on the orientation.
\bigskip

\centerline{\epsfbox{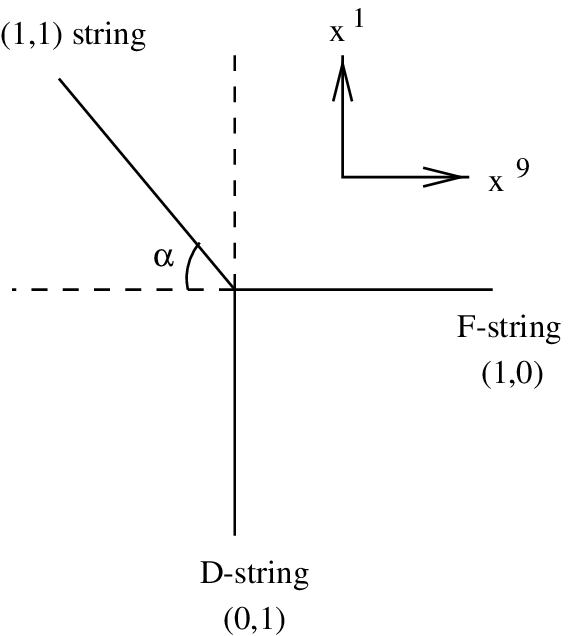}}
\centerline{Fig.2: F-string ending on a D-string, with $\tphi=0$.}
\bigskip

Now, for this simple case, we can see that the force-balance property
holds. The three strings, labelled by their $(p,q)$ charges, have
tensions given by\refs{\schwmult}
\eqn\tensions{
T_{p,q} = \sqrt{p^2 + {q^2\over g^2}}\, T_{1,0}}
From Eqs.\anglealpha\ and \tensions\ it follows that
\eqn\balancecond{
\eqalign{
T_{1,1}\cos\alpha &= T_{1,0}\cr
T_{1,1}\sin\alpha &= T_{0,1} = {1\over g}T_{1,0}}}

Thus, examination of the $U(1)$ gauge theory on the world-sheet of a
D-string has yielded rather nontrivial information. Linear growth of
the scalar potential in 1d translates, by the BPS condition, into a
linearly rising deformation of the D-string on which the point charge
is inserted. The angle is determined by the point charge, and
satisfies the zero-sum condition on the tensions viewed as vectors. 

More general 3-string junctions follow from the one considered above
by application of S-duality transformations. However, so far we have
dealt only with the case where the Ramond-Ramond scalar $\tphi$ is not
excited on the D-string worldsheet. If we make an S-duality, in
general a nonzero RR scalar background will be turned on. Therefore,
we should start by considering the F-D junction (an F-string ending
on a D-string), as above, but in the presence of an arbitrary RR scalar
background. Then S-duality can be used to generate all the other
cases. (It was emphasized in Ref.\refs{\gabzwie} that the most general
3-string junctions are the ones obtained by S-duality on the F-D
junction, at least if one restricts attention to stable and not just
marginally stable states.)

For nonzero $\tphi$, the D-string acquires a Chern-Simons coupling
$\int \tphi\wedge F$ on its worldsheet\refs{\chernsimons}. This shifts
the canonical momentum conjugate to $A_1$, and hence changes the
quantization law by the replacement
\eqn\changequant{
F_{01} \to F_{01} + g\tphi }
in Eq.\gausslaw\ above. As a result, the classical solution
Eq.\simpsol\ for the scalar potential is replaced by:
\eqn\newsol{
\eqalign{
A_0 &= (\tphi - 1)\, g x^1,\quad x^1>0\cr
&= \tphi\, g x^1,\qquad\qquad x^1<0\cr}}
We maintain the BPS nature of the solution by imposing Eq.\xnine.

The interpretation of this solution is as follows. The original
D-string running along $x^1$ has been completely deformed, on both
sides of the inserted charge. The inserted charge is, as before,
associated with an F-string running along the $X^9$ axis. We have thus
obtained the most general F-D junction, and this time (because the RR
scalar is turned on) the F and D strings are not perpendicular. Thus
the physical configuration, for $\tphi\ne 0$, is shown in Fig.3. From
Eqs.\newsol\ and \xnine, the angles $\alpha$ and $\beta$ are given by
\eqn\modangles{
\eqalign{
\tan\alpha &= {1\over g(1-\tphi)}\cr
\tan\beta &= -{1\over g\tphi}\cr}}
\bigskip

\centerline{\epsfbox{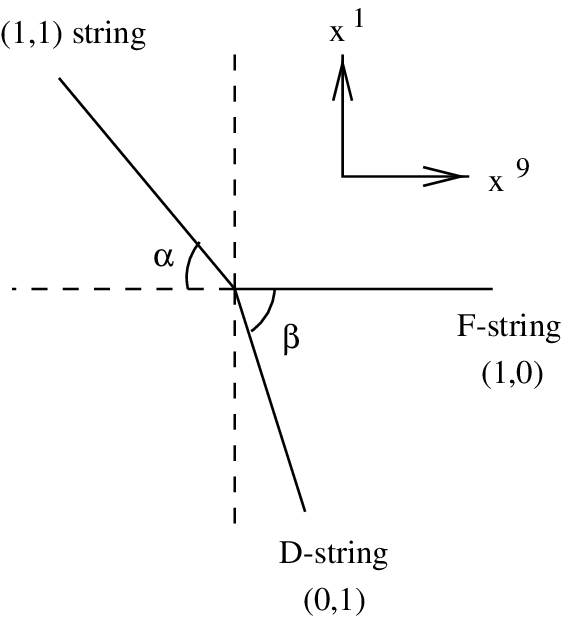}}
\centerline{Fig.3: F-D junction for $\tphi\ne 0$.}
\bigskip

The tension formula Eq.\tensions\ is modified\refs{\schwmult}, in the
presence of $\tphi$, to:
\eqn\modtension{
T_{p,q} = \sqrt{(p-q\tphi)^2 + {q^2\over g^2}}\, T_{1,0}}
It is now again a straightforward matter to check that the force
balance conditions hold:
\eqn\modbalance{
\eqalign{
T_{1,1}\cos\alpha &= T_{1,0} + T_{0,1}\cos\beta\cr
T_{1,1}\sin\alpha &= T_{0,1}\sin\beta }}

With these, we can make an SL(2,Z) transformation to the general
case. Under a transformation with the matrix $\pmatrix{p& q\cr r
&s\cr}$ with $ps-qr=1$, the F-string is mapped to a $(p,r)$ string,
and the D-string to a $(q,s)$ string. The condition $ps-qr=1$ is
precisely the one proposed in Ref.\refs{\gabzwie} to characterize the
allowed 3-string junctions. 

The final configuration after the SL(2,Z) transformation is similar to
that in Fig.3, with the above replacements for the string charges, and
with angles given by
\eqn\finalangles{
\eqalign{
\tan\alpha &= {1\over g\left( (p-r\tphi)( (p+q)-(r+s)\tphi) +
{r^2\over g^2}+{rs\over g^2}\right)}\cr
\tan\beta &= -{1\over g\left((p-r\tphi)(s\tphi-q)-
{rs\over g^2}\right)}\cr}}
Thus we have found the most general BPS saturated, stable 3-string
junction starting from very simple considerations. The tensions and
angles of the junction satisfy the condition, which is intuitively
rather obvious, that the vector sum of the tensions sums to zero. The
stable junctions are those obtained by SL(2,Z) transformation of the
F-D junction. To get all such stable junctions, we needed the F-D
junction in the presence of an arbitrary RR scalar background.

To understand better the nature and role of 3-string junctions is an
important problem. They are known to arise from M-theory by starting
with a 2-brane in a ``pants'' configuration and wrapping each of the
outgoing tubes on a different cycle of some
2-torus\refs{\schwrev}. One may hope to get some insight by studying
the 2-brane worldvolume along the lines of Ref.\refs{\calmal} and the
present paper. 

The 3-string junction could also be studied in the full Born-Infeld
theory\refs{\calmal,\gibbons,\lpt,\thor} rather than the linearized
Maxwell case as we have done. The proposal of Ref.\refs{\schwrev} to
look for them as solutions of supergravity equations, remains to be
carried out. It would be interesting to see if 3-string junctions can
be recovered in the matrix approach to string
theory\refs{\banks}. Finally, it remains to investigate the
proposal\refs{\gabzwie} that multi-pronged strings could play a
fundamental role in nonperturbative string field theory.

\bigskip

\noindent{\bf Acknowledgements:} We are grateful to Atish Dabholkar,
Abhishek Dhar, Ashoke Sen and Barton Zwiebach for useful discussions.

\listrefs    

\end